\documentclass[journal=apchd5,manuscript=article, layout=twocolumn]{achemso}

\usepackage[T1]{fontenc}
\usepackage[charter]{mathdesign}\usepackage{XCharter} 
\makeatletter\AtBeginDocument{\acs@layout@ten}\makeatother 
\usepackage{siunitx} 
\usepackage{booktabs} 
\usepackage{multirow} 
\usepackage[switch]{lineno} 

\title{Cryogenic Luminescence Thermometry of Upconversion Micro- and Nanoparticles via Deep Learning}

\author{Pablo José Coria Ledo}
\affiliation{Universidad de Buenos Aires, Facultad de Ciencias Exactas y Naturales, Departamento de F\'{\i}sica. Buenos Aires, Argentina}
\author{Lina W\"olcken}
\affiliation{Universidad de Buenos Aires, Facultad de Ciencias Exactas y Naturales, Departamento de F\'{\i}sica. Buenos Aires, Argentina}
\author{Lucas R. Mendicino}
\affiliation{Universidad de Buenos Aires, Facultad de Ciencias Exactas y Naturales, Departamento de F\'{\i}sica. Buenos Aires, Argentina}
\alsoaffiliation{CONICET - Universidad de Buenos Aires, Instituto de F\'{\i}sica de Buenos Aires (IFIBA). Buenos Aires, Argentina}
\author{Santiago Gliosca}
\affiliation{Universidad de Buenos Aires, Facultad de Ciencias Exactas y Naturales, Departamento de F\'{\i}sica. Buenos Aires, Argentina}
\alsoaffiliation{CONICET - Universidad de Buenos Aires, Instituto de F\'{\i}sica de Buenos Aires (IFIBA). Buenos Aires, Argentina}
\author{M. Claudia Marchi}
\affiliation{Universidad de Buenos Aires, Facultad de Ciencias Exactas y Naturales, Centro de Microscop\'{\i}as Avanzadas (CMA). Buenos Aires, Argentina}
\alsoaffiliation{CONICET - Universidad de Buenos Aires, Instituto de F\'{\i}sica de Buenos Aires (IFIBA). Buenos Aires, Argentina}
\author{Marcos Illescas}
\affiliation{Universidad de Buenos Aires, Facultad de Ciencias Exactas y Naturales, Departamento de Qu\'{\i}mica Inorg\'anica, Anal\'{\i}tica y Qu\'{\i}mica F\'{\i}sica. Buenos Aires, Argentina}
\alsoaffiliation{CONICET - Universidad de Buenos Aires, Instituto de Qu\'{\i}mica F\'{\i}sica de los Materiales, Medio Ambiente y Energ\'{\i}a (INQUIMAE). Buenos Aires, Argentina}
\author{Beatriz Barja}
\affiliation{Universidad de Buenos Aires, Facultad de Ciencias Exactas y Naturales, Departamento de Qu\'{\i}mica Inorg\'anica, Anal\'{\i}tica y Qu\'{\i}mica F\'{\i}sica. Buenos Aires, Argentina}
\alsoaffiliation{CONICET - Universidad de Buenos Aires, Instituto de Qu\'{\i}mica F\'{\i}sica de los Materiales, Medio Ambiente y Energ\'{\i}a (INQUIMAE). Buenos Aires, Argentina}
\author{Christian Tom\'as Schmiegelow}
\affiliation{Universidad de Buenos Aires, Facultad de Ciencias Exactas y Naturales, Departamento de F\'{\i}sica. Buenos Aires, Argentina}
\alsoaffiliation{CONICET - Universidad de Buenos Aires, Instituto de F\'{\i}sica de Buenos Aires (IFIBA). Buenos Aires, Argentina}
\email{schmiegelow@df.uba.ar}

\begin{document}

\begin{abstract}
Luminescence thermometry with single or optically trapped particles is hampered by weak signals that fluctuate with alignment, collection efficiency, and excitation conditions and, for $\text{Er}^{3+}$-based probes, by the loss of sensitivity of the luminescence intensity ratio (LIR) at cryogenic temperatures. Here we use a one-dimensional convolutional neural network (1D-CNN) to predict the temperature directly from single emission spectra of $\text{NaYF}_4\text{:Yb}^{3+}\text{,Er}^{3+}$ micro- and nanoparticles between \SI{-150}{\celsius} and \SI{105}{\celsius}, and compare it with the standard LIR under adverse conditions deliberately reproduced with particle ensembles. From \SI{-150}{\celsius} to \SI{30}{\celsius}, the 1D-CNN reduces the single-spectrum mean absolute error from \SI{5.4}{\celsius} to \SI{1.0}{\celsius} and keeps its temperature uncertainty nearly constant down to the lowest temperatures, where Stark-component features remain temperature dependent while the LIR loses its sensitivity. Under deliberate optical misalignment, its error increases only from \SI{0.58}{\celsius} to \SI{0.88}{\celsius}, whereas that of the LIR rises from \SI{1.4}{\celsius} to \SI{3.5}{\celsius}. The same approach reaches sub-degree accuracy (\SI{0.21}{\celsius}) over a narrow temperature window and a single calibration valid across different pump powers and wavelengths. These results establish 1D-CNNs as robust single-shot thermometers and a step toward fast thermometry of single and trapped particles.
\end{abstract}

\section{Introduction}
Luminescence thermometry exploits the temperature dependence of the emission of luminescent materials to measure temperature remotely, with high spatial resolution and in environments inaccessible to contact probes~\cite{brites2012thermometry,jaque2012luminescence}. Lanthanide-doped upconversion particles are among the most widely used probes~\cite{vetrone2010temperature,brites2019lanthanide}. A standard readout is the luminescence intensity ratio (LIR) between two emission bands connected to thermally coupled excited states, whose relative populations are commonly described by a Boltzmann distribution~\cite{wade2003fluorescence,suta2020theoretical}. The LIR method is simple, physically grounded, and self-referenced, which has made it the method of choice in most applications. In this study, we use $\text{NaYF}_4\text{:Yb}^{3+}\text{,Er}^{3+}$ particles and define the LIR from the green emission bands associated with the $^2\text{H}_{11/2}$ and $^4\text{S}_{3/2}$ manifolds of $\text{Er}^{3+}$~\cite{vetrone2010temperature,brites2019lanthanide,jandar2025improved,debasu2013allinone,brites2019self,kilbane2016farfield,hyppanen2017environmental,geitenbeek2017nayf4}.

The LIR method, however, relies on a small fraction of the emission spectrum and requires spectra with a good signal-to-noise ratio. This becomes a limitation in some of the most demanding scenarios. When single particles are used as local thermometers, or when particles are optically trapped to study their laser cooling or heating~\cite{maestro2025intrinsic,rahman2017laser,luntzmartin2021laser}, the collected signal is weak and fluctuates with the alignment, the collection efficiency, and the excitation power, so that long integration times or averaging over many spectra are needed to reach an acceptable temperature uncertainty~\cite{rodriguezsevilla2018temperature}. In addition, at cryogenic temperatures the population of the upper $^2\text{H}_{11/2}$ level becomes negligible and the LIR loses its sensitivity, although other spectral features, such as the intensities and widths of individual Stark components, remain temperature dependent~\cite{suta2020theoretical,possmayer2026boltzmann}. Finally, the emission of individual particles can differ from that of the ensemble used for calibration, which limits the transfer of ensemble calibrations to single-particle thermometry~\cite{pickel2018apparent,zharkov2023upconversion}.

Machine learning offers a way to exploit the full spectral information without an explicit physical model. Principal component analysis, support vector machines, and neural networks have been applied to luminescence thermometry, generally outperforming the LIR~\cite{gavrilovic2024luminescence,xu2024luminescence,polikovskiy2025combining,cui2023convolutional,xu2023convolutional,ciric2026spectrasensml}; among them, neural networks have often yielded the best performance~\cite{polikovskiy2025combining}. These studies, however, have mainly addressed accuracy under well-controlled acquisition conditions, often with materials specifically designed for thermometry~\cite{polikovskiy2025combining}, and robustness has been examined mainly against specific perturbations such as background light or tissue-induced spectral distortions~\cite{xu2024luminescence,xu2023convolutional,xu2025ultrarobust}. Here we assess whether a neural network can sustain its performance under the weak and fluctuating signals relevant to single and trapped particles, and at cryogenic temperatures where the LIR loses its sensitivity. Since the LIR remains the reference method in the field, we use it as the baseline for this assessment.

In this work, we use a one-dimensional convolutional neural network (1D-CNN) to predict the temperature directly from single emission spectra of $\text{NaYF}_4\text{:Yb}^{3+}\text{,Er}^{3+}$ micro- and nanoparticles between \SI{-150}{\celsius} and \SI{105}{\celsius}. Using concentrated particle samples, we deliberately reproduce the adverse conditions expected for single and trapped particles (low signal, cryogenic temperatures, optical misalignment, and changes in excitation power and wavelength) and evaluate the thermometer in each case. Compared with the LIR, the 1D-CNN reduces the single-spectrum error by about a factor of five, keeps its temperature uncertainty nearly constant down to the lowest temperatures, and remains accurate under deliberate optical misalignment. It also reaches sub-degree accuracy when calibrated over a narrow temperature window and can be calibrated simultaneously for different excitation conditions. These results establish the 1D-CNN as a robust single-shot thermometer and a step toward fast thermometry of single and optically trapped particles.

\begin{figure*}[!ht]
\centering
\includegraphics[width=\linewidth]{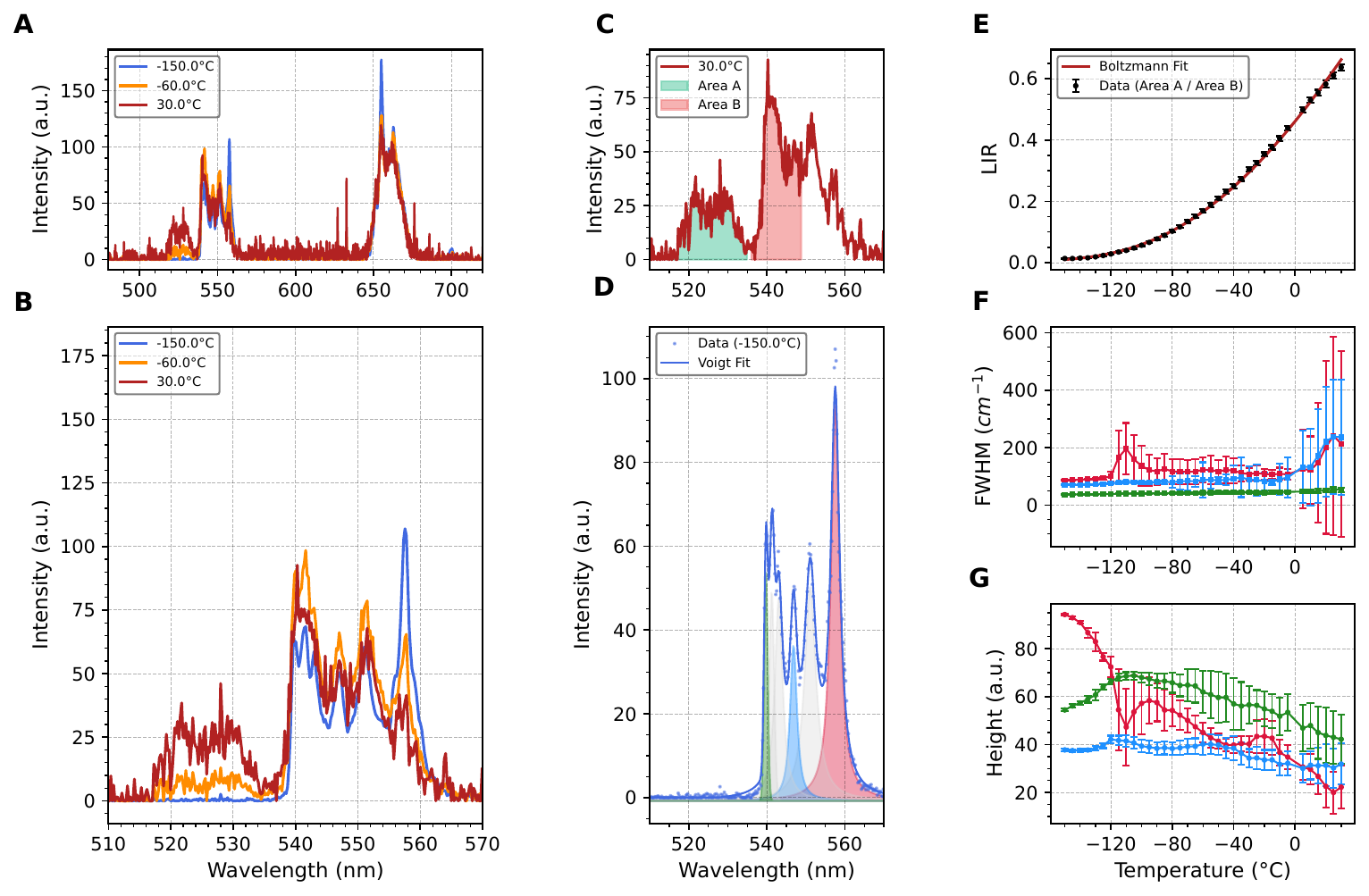}
\caption{Emission analysis and LIR thermometric calibration of $\text{NaYF}_4\text{:Yb}^{3+}\text{,Er}^{3+}$ particles (Sample A). \textbf{(A)} Luminescence spectra in the visible region (480--720~\SI{}{\nano \meter}) recorded at three representative sample temperatures. \textbf{(B)} Magnified view of the spectra in the green emission region (510--570~\SI{}{\nano \meter}) illustrating the thermal evolution of the profiles. \textbf{(C)} Schematic of the integration regions used for the ratiometric method on the spectrum at the highest temperature (\SI{30}{\celsius}). Area A (515--535~\SI{}{\nano \meter}) and Area B (536--549~\SI{}{\nano \meter}) correspond to the numerator and denominator for the LIR calculation, respectively. \textbf{(D)} Deconvolution of the spectral peaks at the minimum temperature using multiple Voigt profiles. Raw data (gray dots), global fit (solid blue line), and individual components are shown, highlighting three specific Stark components in color for tracking. \textbf{(E)} LIR thermometric calibration curve fitted using a Boltzmann physical model with a background offset parameter. \textbf{(F)} Evolution of the full width at half maximum (FWHM) expressed in \SI{}{\per \centi \meter} as a function of temperature for the three highlighted Stark peaks. \textbf{(G)} Thermal dependence of the peak intensity (height) for the same Stark components, reflecting the thermal redistribution of population. For panels (E), (F), and (G), data points represent mean values, and error bars indicate their respective standard deviations. Prior to display in panels (A), (B) and (C), and to the Voigt deconvolution fitting in panel (D), the spectra were normalized using an $L^2$ (Euclidean) norm.}
\label{fig:panel}
\end{figure*}

\section{Experimental Procedure}

\subsection{Sample Preparation and Optical Setup}
Two samples of $\text{NaYF}_4\text{:Yb}^{3+}\text{,Er}^{3+}$ particles were used in this study. \textbf{Sample A} consisted of larger, spherical particles with a diameter of approximately \SI{1}{\micro\meter}. \textbf{Sample B} consisted of smaller nanoparticles with a disk-like morphology (approximated diameter $\sim$\SI{100}{\nano\meter} and thickness $\sim$\SI{55}{\nano\meter}). Both samples were synthesized at INQUIMAE by thermal decomposition of sodium and lanthanide trifluoroacetates dissolved in a mixture of oleic acid and 1-octadecene~\cite{ye2010morphologically}, with nominal doping of 20 mol\% $\text{Yb}^{3+}$ and 2 mol\% $\text{Er}^{3+}$ in the $\beta$-phase host; sizes and morphology were determined by scanning electron microscopy (SEM).

For sample preparation, concentrated colloidal dispersions of these particles in hexane were deposited onto \SI{0.6}{\milli\meter} thick aluminum substrates. The deposition was performed via drop-casting in \SI{10}{\micro\litre} increments, allowing the solvent to fully evaporate between successive drops, until a uniform white coating covering an area of approximately \SI{8}{\milli\meter} $\times$ \SI{8}{\milli\meter} was achieved.

The thermal characterization spanned a range from \SI{-150}{\celsius} to \SI{105}{\celsius} using two complementary setups: from \SI{-150}{\celsius} to \SI{30}{\celsius}, the particles were placed inside an optical cryostat equipped with a PID-controlled resistive heater, whereas from room temperature up to \SI{105}{\celsius}, an analogous setup was used outside the cryostat with a Peltier element as heater. Excitation was provided by a Ti:Sa laser; incident power densities were estimated from the total beam power and the estimated beam size at the focus of the lens. The resulting luminescence was directed into a commercial CCD spectrometer (Thorlabs CCS200) with a spectral resolution better than \SI{2}{\nano\meter} (FWHM). Representative emission spectra at three temperatures across the cryostat range are shown in Fig.~\ref{fig:panel}(A).

\subsection{Data Acquisition}
Acquisition conditions were varied throughout the study to encompass a wide range of experimental conditions. Temperature sweeps employed both fine and coarse steps, with between 100 and 500 spectra acquired at each setpoint depending on the specific case. Furthermore, the excitation parameters were varied across trials, using different excitation wavelengths and incident power densities, with the latter ranging from \SI{5}{\watt\per\centi\meter\squared} to \SI{300}{\watt\per\centi\meter\squared}. Throughout this work, the reported temperatures correspond to the substrate setpoint; at the higher excitation power densities, laser-induced heating may raise the particle temperature slightly above this value, by an amount that depends on the power density~\cite{pickel2018apparent}. Additionally, one dataset was acquired by intentionally introducing variable optical misalignment to emulate adverse measurement scenarios, while all other spectra were recorded under optimal alignment to yield a maximum signal-to-noise ratio. Finally, all acquired spectra were normalized by their respective integration times.

To process this spectral data and evaluate the thermometric performance of the particles, two approaches were implemented and compared: the classical Luminescence Intensity Ratio (LIR) and a one-dimensional convolutional neural network (1D-CNN).

\subsection{LIR Thermometry}
\label{subsec:lir}

The green emission bands of $\text{Er}^{3+}$ are widely used for ratiometric thermometry~\cite{vetrone2010temperature,geitenbeek2017nayf4}. Specifically, the LIR is calculated as the ratio of the integrated luminescence intensities corresponding to the decays from the thermally coupled $^2\text{H}_{11/2}$ and $^4\text{S}_{3/2}$ levels to the $^4\text{I}_{15/2}$ ground state. Fig.~\ref{fig:panel}(B) shows a close-up of the green region of the emission spectrum, illustrating the relative intensity variation between these two bands as a function of temperature. As expected, emission from the higher-energy level increases at elevated temperatures as its thermal population grows, whereas emission from the lower-energy level dominates at lower temperatures due to the thermal depopulation of the upper state.

In this work, the LIR was determined by integrating the spectral intensity between \SI{515}{\nano \meter} and \SI{535}{\nano \meter} for the numerator, and between \SI{536}{\nano \meter} and \SI{549}{\nano \meter} for the denominator. This ratio was computed for each individual spectrum at a given temperature. The mean values were then used to calibrate the thermometer by fitting the data to a Boltzmann distribution model (Fig.~\ref{fig:panel}(E)), as shown in Eq.~\ref{boltzmann}:

\begin{equation}
    \text{LIR}(T) = \frac{I_{515\text{--}535}}{I_{536\text{--}549}} = A \exp\left(- \frac{\Delta E}{k_B T}\right) + B
    \label{boltzmann}
\end{equation}

where $A$ is a material-dependent constant governed by the degeneracy of the thermally coupled levels, the spontaneous emission rates, the photon energies of each transition, and the quantum efficiency; $\Delta E$ is the effective energy gap between the thermally coupled levels; $k_B$ is the Boltzmann constant; $T$ is the absolute temperature of the sample; and $B$ is an offset constant. The parameters were fitted independently for each dataset, and the resulting values of $\Delta E$ are reported in the corresponding case studies for comparison with values reported for the $^2\text{H}_{11/2}$--$^4\text{S}_{3/2}$ gap in the literature~\cite{geitenbeek2017nayf4}. 

A major weakness of classical ratiometric methods is that they use only a small fraction of the available spectral information to extract the temperature. By solely considering the ratio of the integrated emission intensities of two bands, these methods restrict the analysis to a limited spectral region and ignore individual emission peak features, such as width and height, which often exhibit a strong temperature dependence. To illustrate this, a spectral deconvolution of the peaks between \SI{536}{\nano \meter} and \SI{570}{\nano \meter} was performed, as shown in Fig.~\ref{fig:panel}(D). From these fits, three peaks were selected to track their full width at half maximum (FWHM) and peak height (intensity) as a function of temperature. The results, presented in Fig.~\ref{fig:panel}(F)--(G), demonstrate that both peak width and height exhibit distinct temperature dependencies that could be exploited to develop more accurate thermometers. Notably, several of these features, such as the heights of the selected Stark components (Fig.~\ref{fig:panel}(G)), continue to vary markedly below \SI{-100}{\celsius}, where the $^2\text{H}_{11/2}$ emission, and with it the sensitivity of the LIR, has essentially vanished. However, formulating a comprehensive physical model to describe the thermal evolution of all these spectral features simultaneously is highly complex, and routine spectral deconvolution can be computationally expensive. Machine learning methods offer an alternative approach to exploit the full potential of the spectral morphology for temperature prediction. Specifically, this work employs a one-dimensional convolutional neural network (1D-CNN), a deep learning algorithm well suited to sequential data such as spectra~\cite{kiranyaz2021onedimensional}.

\subsection{1D-CNN Thermometry}
For the training and evaluation of the model, the spectra of each dataset were randomly divided into training (\SI{90}{\percent}) and test (\SI{10}{\percent}) subsets at the level of individual spectra, stratified by temperature setpoint. Training and test spectra therefore share the same set of temperature setpoints and acquisition runs, and the reported metrics quantify single-spectrum performance at calibrated temperatures. Interpolation to temperatures absent from the training set is tested separately in Section~\ref{subsec:misalignment}. Each spectrum was wavelength-aligned by linear interpolation onto a fixed common axis spanning 450--750~nm with 1313 uniformly spaced points. Prior to input to the network, each interpolated spectrum was scaled to unit norm using an $L^2$ (Euclidean) normalizer, which suppresses global intensity fluctuations and makes the model rely on temperature-dependent spectral morphology (relative band amplitudes and linewidths). The network was trained on temperatures in degrees Celsius, whereas the Boltzmann fit (Eq.~\ref{boltzmann}) was performed in kelvin.

A 1D-CNN regressor was implemented in Python using TensorFlow/Keras~\cite{abadi2016tensorflow,chollet2015keras}. The feature extractor consisted of two convolutional blocks: Conv1D (48 filters, kernel size 5, ReLU) + batch normalization~\cite{ioffe2015batch} + max pooling (pool size 2), followed by a second Conv1D block (32 filters, kernel size 5, ReLU) + batch normalization + max pooling. The regression head consisted of a flatten layer followed by a dense layer with 192 ReLU units and a dropout layer~\cite{srivastava2014dropout} (rate 0.20), and a final linear neuron for continuous temperature output.

The network was trained for up to 200 epochs with a batch size of 64 using the Adam optimizer~\cite{kingma2014adam} (learning rate $1.4\times10^{-4}$) to minimize the mean squared error (MSE), while monitoring the root mean squared error (RMSE) as an additional metric. Training used early stopping on the loss of a validation subset held out from the training data, with a patience of 20 epochs (restoring the best weights) and a learning-rate scheduler (\texttt{ReduceLROnPlateau}) that reduced the learning rate by a factor of 0.5 after 5 epochs without improvement (minimum learning rate $10^{-7}$).

To select the model hyperparameters, a separate optimization stage was performed using KerasTuner~\cite{omalley2019kerastuner} (random search~\cite{bergstra2012random}) on the same preprocessed inputs (integration-time normalization, interpolation to the common wavelength grid, and $L^2$ normalization). The search objective was the validation MSE. A total of 30 random hyperparameter configurations were evaluated, and each configuration was trained twice to reduce stochastic variability. Each trial used up to 50 epochs (batch size 32) with early stopping on the validation loss (patience 10, restoring best weights). The explored search space comprised the number of filters in each convolutional layer (16--64 and 32--128), kernel sizes (3--7 and 3--5), dense-layer width (64--256), dropout rate (0.2--0.5), and the Adam learning rate sampled logarithmically between $10^{-4}$ and $10^{-2}$. The final architecture and training settings reported above correspond to the best-performing hyperparameter set identified by this procedure. The optimization was performed once, on the Sample B dataset (Section~\ref{subsec:misalignment}), and the resulting architecture was used unchanged for all case studies.

From a practical standpoint, the computational cost of the 1D-CNN is concentrated in the training phase. Once trained, the model occupies less than \SI{8}{\mega\byte} and predicts the temperature from a single spectrum in approximately \SI{12}{\milli\second} on a standard laptop CPU (Intel Core i7-1255U, without GPU acceleration), compared with \SI{0.02}{\milli\second} for the LIR. Although three orders of magnitude slower, this latency is well within the requirements of most real-time monitoring applications and compatible with deployment on low-cost embedded hardware. Moreover, the effective acquisition time of the LIR increases with the number of spectra that must be averaged to reach a comparable uncertainty (Section~\ref{subsec:misalignment}).

\section{Case Studies}

This section evaluates the performance of the thermometers calibrated under the various experimental conditions investigated. In all case studies, performance is quantified on the held-out test spectra using the prediction residual $T_{\mathrm{pred}}-T_{\mathrm{ref}}$. The aggregate figures of merit used throughout are the mean absolute error (MAE), which captures the typical single-spectrum deviation; the root mean squared error (RMSE), which penalizes occasional large errors more strongly; and the maximum absolute error, which provides a conservative bound on worst-case performance within the sampled conditions. For the first two scenarios, both thermometric approaches are compared directly. Subsequently, having shown the advantages of the machine learning approach in these two scenarios, the 1D-CNN's performance is explored under specific conditions. Together, these results establish the 1D-CNN as a robust and versatile alternative to conventional ratiometric thermometry in these systems.

\subsection{Cryogenic Temperature Range}
\label{subsec:cryo}
The first case study evaluated the emission spectra of Sample A over a broad thermal range, spanning from cryogenic conditions (\SI{-150}{\celsius}) to room temperature (\SI{30}{\celsius}) in \SI{5}{\celsius} increments. Particle excitation was performed using a constant power density of approximately \SI{70}{\watt\per\centi\meter\squared} at a wavelength of \SI{976}{\nano\meter}. Approximately 250 spectra were acquired at each temperature setpoint to calibrate two distinct thermometers: one relying on the LIR fitted to a Boltzmann distribution, and another employing a 1D-CNN to predict the temperature directly from the full emission spectrum. For both methods, the same 90:10 split described above was used, with the LIR calibrated on the training subset.

\begin{figure}[!h]
\centering
\includegraphics[width=\linewidth]{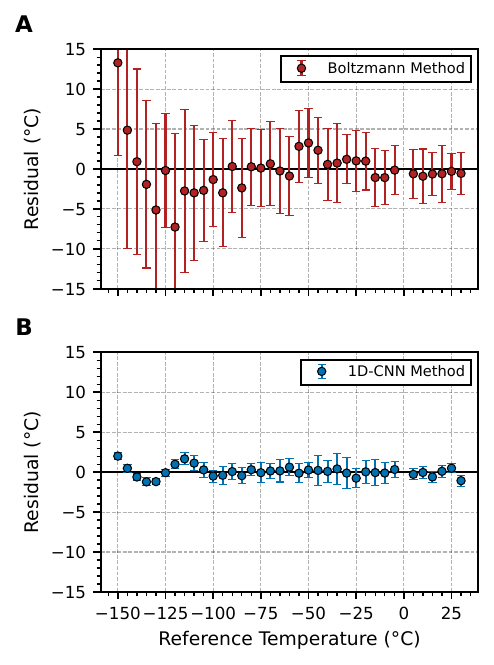}
\caption{Comparison of temperature-dependent residuals for \textbf{(A)} the LIR and \textbf{(B)} the 1D-CNN thermometers. Data points represent the mean residual of the test set predictions at each temperature setpoint, with error bars indicating the standard deviation of the residuals at that setpoint.}
\label{fig:residuals_cerebritos_crio}
\end{figure}

The LIR was calibrated on setpoint-averaged ratios as described in Section~\ref{subsec:lir}, yielding $\Delta E = 661 \pm 6$~\si{\per\centi\meter}, whereas both thermometers were evaluated on individual test spectra.

Once the LIR thermometer was calibrated, it was employed to predict the temperature across the dedicated test
set. Overall, the classical method yielded an MAE of \SI{5.40}{\celsius}, an RMSE of \SI{7.47}{\celsius}, and a maximum prediction error exceeding \SI{33}{\celsius}.

The same evaluation was performed on the neural network-based thermometer, yielding an MAE of \SI{1.00}{\celsius}, an RMSE of \SI{1.27}{\celsius}, and a maximum error of \SI{4.66}{\celsius}. The 1D-CNN thus reduces the single-spectrum errors by a factor of about five.

Beyond this overall improvement, Fig.~\ref{fig:residuals_cerebritos_crio} displays the residuals of both thermometers as a function of temperature. While the performance of both methods deteriorates at the lowest studied temperatures, this degradation is significantly less severe for the 1D-CNN. This is primarily because, at cryogenic temperatures, the intensity of the emission band corresponding to the LIR numerator approaches the noise level of the detector, causing the ratiometric calculation to lose significant thermal sensitivity. Although the loss of a highly temperature-sensitive spectral feature is also detrimental to the neural network, it is not as severely impacted as the classical thermometer, which relies entirely on this single parameter, because the 1D-CNN employs multiple spectral features simultaneously to make its predictions. This is reflected in the temperature uncertainty $\delta T$, taken as the standard deviation of single-spectrum predictions at each setpoint (error bars in Fig.~\ref{fig:residuals_cerebritos_crio}): for the LIR it grows from  ~5 to ~20\si{\celsius} towards the lowest temperatures, as its relative sensitivity drops, whereas for the 1D-CNN it remains approximately constant at $\lesssim$2\si{\celsius}. The effective sensitivity of the 1D-CNN is thus sustained at low temperature by spectral features, such as the Stark-component heights in Fig.~\ref{fig:panel}(G), that remain temperature dependent where the LIR does not. 

A closer inspection of the 1D-CNN predictions in Fig.~\ref{fig:residuals_cerebritos_crio}, however, reveals a slight periodic structure in the residuals near the lowest temperatures of the sweep. We attribute this minor artifact to insufficient training data in this challenging thermal regime. This is supported by the results of Section~\ref{subsec:misalignment}: in a comparable measurement on a different sample (Sample B), with twice as many spectra per setpoint and a higher signal-to-noise ratio, no such structure is observed (Fig.~\ref{fig:residuals}(B)). Nevertheless, the deep learning-based thermometer not only outperforms the classical method in overall metrics, but also serves as a more reliable tool for temperature predictions in these systems at cryogenic temperatures, thereby mitigating a fundamental physical limitation of the LIR method.

\subsection{Impact of Optical Misalignment}
\label{subsec:misalignment}

Another case study evaluated the impact of optical misalignment during spectral acquisition on the performance of the thermometric methods. To address the minor artifacts observed at cryogenic temperatures in the previous section and to facilitate a more intensive training phase for the neural network, the experimental conditions were intentionally modified. In this case, Sample B was used, and excitation was provided at a substantially higher constant power density of approximately \SI{300}{\watt\per\centi\meter\squared} with a wavelength of \SI{976}{\nano\meter}. This higher power density was chosen to substantially improve the signal-to-noise ratio of each individual spectrum. The thermal sweep spanned from \SI{-145}{\celsius} to \SI{30}{\celsius} in \SI{5}{\celsius} increments and was conducted twice under distinct optical alignment conditions. The first sweep was performed under optimal alignment to achieve the highest possible signal-to-noise ratio. Subsequently, the procedure was repeated while intentionally introducing random manual misalignments to the optical elements during spectral acquisition. Additionally, the cryostat was mechanically translated to acquire spectra from different regions of the sample. In both scenarios, approximately 500 individual spectra were recorded at each temperature setpoint (roughly double the amount acquired in the previous case), providing a much larger dataset to train the model more thoroughly. The LIR was calibrated on the setpoint-averaged ratios of the aligned training spectra only, whereas the 1D-CNN was trained on both aligned and misaligned training spectra; both thermometers were tested on the same held-out spectra from both conditions.

\begin{figure}[!h]
\centering
\includegraphics[width=\linewidth]{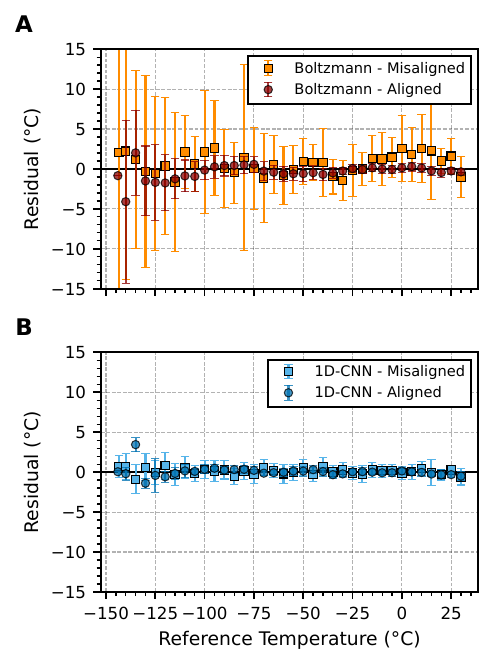}
\caption{Temperature-dependent residuals for \textbf{(A)} the LIR and \textbf{(B)} the 1D-CNN thermometric methods under optimal alignment and deliberate optical misalignment. For each temperature setpoint, the data points denote the average residual computed from the test set, while the error bars represent the standard deviation of the residuals at that setpoint.}
\label{fig:residuals}
\end{figure}

The Boltzmann fit of the aligned calibration data yielded $\Delta E = 754 \pm 3$~\si{\per\centi\meter}. To evaluate the robustness against these adverse conditions, the performance metrics of both thermometers were analyzed separately across the aligned and misaligned datasets, with the temperature-dependent residuals for both scenarios presented in Fig.~\ref{fig:residuals}. As observed in Fig.~\ref{fig:residuals}(A), the classical LIR method exhibited severe degradation under suboptimal alignment, with its MAE deteriorating from \SI{1.36}{\celsius} to \SI{3.54}{\celsius} (a \SI{160}{\percent} increase) and its RMSE worsening from \SI{2.71}{\celsius} to \SI{6.05}{\celsius}. It is worth noting, however, that under optimal alignment and within a restricted, non-cryogenic temperature range, the classical LIR method performs remarkably well, achieving an accuracy comparable to that of the neural network. Yet, this performance is confined to well-controlled experimental conditions and a limited thermal window. As previously noted, the thermal depopulation of the upper energy level at cryogenic temperatures inherently increases prediction uncertainty. While this physical loss of sensitivity broadens the residual spread for both thermometers at the lower end of the thermal sweep, the LIR approach is disproportionately impaired, causing its maximum prediction error to exceed \SI{56}{\celsius}. 

In contrast, the 1D-CNN residuals depicted in Fig.~\ref{fig:residuals}(B) demonstrate stable performance across both alignment scenarios. The network's MAE experienced only a slight increase from \SI{0.58}{\celsius} to \SI{0.88}{\celsius}, while its RMSE rose moderately from \SI{0.90}{\celsius} to \SI{1.18}{\celsius}, with a maximum prediction error restricted to \SI{7.89}{\celsius}. The benefits of the intensified training and improved signal-to-noise ratio are reflected in these overall metrics, which improved substantially for both methods compared to the previous case study. Notably, the model trained on this larger dataset, which includes both aligned and misaligned spectra, outperformed the model of the previous section even on its misaligned test subset (MAE of \SI{0.88}{\celsius} vs.\ \SI{1.00}{\celsius}), consistent with the benefit of a larger training set discussed in Section~\ref{subsec:cryo}. Furthermore, no clear periodic structure is observed in the cryogenic residuals, even in the misaligned data. These results confirm that a larger, higher-quality training dataset removes such artifacts. More importantly, the 1D-CNN reaches this accuracy from single spectra, whereas LIR thermometry usually relies on averaging many spectra to improve the signal-to-noise ratio. Given the larger spread of the single-spectrum LIR residuals (Fig.~\ref{fig:residuals}(A)), the LIR would require averaging of the order of tens of spectra to reach a comparable uncertainty, at the cost of time resolution. A single 1D-CNN calibration covering both alignment conditions therefore provides reliable single-shot readout under suboptimal optical collection.

To verify that the trained 1D-CNN acts as a genuine thermometer (continuous regressor) rather than a discrete temperature classifier, an additional interpolation test was performed using the aligned Sample B dataset~\cite{ciric2026spectrasensml}. Specifically, the network was trained on \SI{90}{\percent} of the spectra acquired every \SI{10}{\celsius} and then evaluated on a test set comprising all spectra at the intermediate temperatures (offset by \SI{5}{\celsius}), which were never seen during training, together with the remaining \SI{10}{\percent} of held-out spectra at the trained setpoints. On this combined test set, the model maintains good accuracy, yielding an MAE of \SI{0.80}{\celsius}, an RMSE of \SI{1.22}{\celsius}, and a maximum absolute error of \SI{7.13}{\celsius}. This confirms that the 1D-CNN learns a continuous mapping between spectral morphology and temperature, rather than assigning spectra to a finite set of trained temperature labels.

\subsection{Extended Temperature Range}

Having established the 1D-CNN as a robust tool for developing thermometers from luminescence spectra, subsequent studies focused on specific cases of interest to evaluate the machine learning model's performance under various experimental conditions. 

The first test aimed to extend the operational range of the thermometer beyond room temperature. For this purpose, the luminescence spectra of Sample A were measured as a function of temperature from ambient conditions up to \SI{105}{\celsius} in \SI{5}{\celsius} increments. These measurements were conducted using an experimental setup analogous to the cryostat configuration, but operating outside the cryostat with a Peltier element as heater. Approximately 150 spectra were acquired at each temperature setpoint under a constant excitation power density of roughly \SI{300}{\watt\per\centi\meter\squared} at a wavelength of \SI{976}{\nano\meter}. The training and testing protocols for the model were the same as in the previous cases. However, the overall dataset employed for this evaluation was constructed by merging the existing cryogenic-to-room-temperature data for these specific particles with the newly recorded measurements above room temperature.

\begin{figure*}[!ht]
\centering
\includegraphics[width=\linewidth]{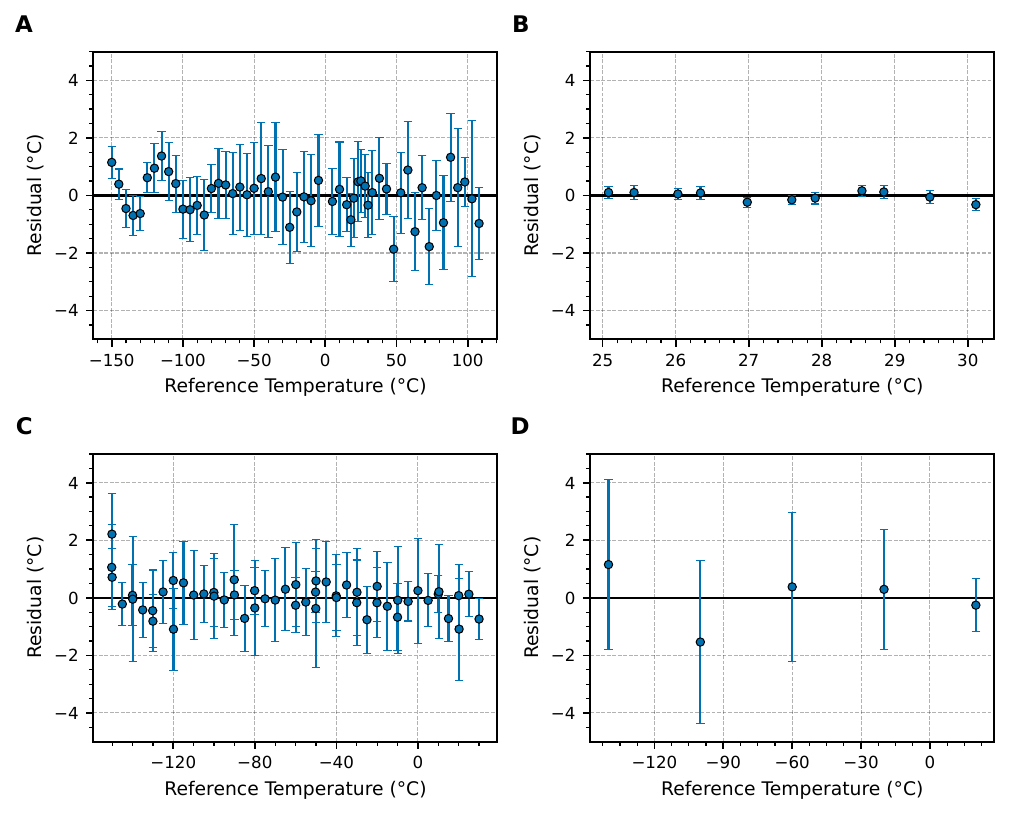}
\caption{Temperature-dependent residuals of the 1D-CNN predictions across various experimental case studies. \textbf{(A)} Extended operational temperature range. \textbf{(B)} Narrow temperature window evaluated with finer thermal intervals. \textbf{(C)} Evaluation under different excitation pump power densities. \textbf{(D)} Performance across varying excitation wavelengths. In all sub-panels, the data points denote the average residual computed from the test set for each temperature setpoint, while the error bars represent the standard deviation of the residuals at that setpoint.}
\label{fig:residuals_panel}
\end{figure*}

The performance of the 1D-CNN in this scenario can be evaluated through its temperature-dependent residuals, presented in Fig.~\ref{fig:residuals_panel}(A). The neural network successfully predicts the temperature with relatively uniform accuracy across the entire investigated thermal range. Specifically, this extended calibration yielded an MAE of \SI{1.12}{\celsius}, an RMSE of \SI{1.43}{\celsius}, and a maximum error of \SI{4.85}{\celsius}. These metrics represent marginal increases of only \SI{12}{\percent}, \SI{13}{\percent}, and \SI{4}{\percent}, respectively, when compared to the results obtained from the narrower cryogenic dataset. This slight increase in prediction error is reasonable, considering that it enabled an extension of the thermometer's upper operational limit by more than \SI{70}{\celsius}, ultimately yielding an effective thermometric tool capable of operating over a total span exceeding \SI{250}{\celsius}. This demonstrates the neural network's capability to predict the temperature of these particles over a broad thermal range. Consequently, it emerges as a promising tool for temperature monitoring in scenarios spanning from cryogenic to above-ambient temperatures, achieved at the cost of only a minor reduction in accuracy relative to models trained on narrower temperature windows.

\subsection{Narrow Temperature Range}

Having observed that extending the operational range of the neural network-based thermometer is possible, albeit at the expense of a slight reduction in prediction accuracy, the question arises as to whether substantially higher accuracy can be achieved by training the model over a narrower temperature range with finer acquisition steps. To address this, luminescence spectra of Sample A were acquired at intervals of approximately \SI{0.5}{\celsius} from \SI{25}{\celsius} to \SI{30}{\celsius}. This experiment used the cryostat setup, maintaining a constant excitation power density of roughly \SI{70}{\watt\per\centi\meter\squared} at a wavelength of \SI{976}{\nano\meter}. Approximately 250 spectra were recorded at each temperature setpoint. 

The model was trained and tested as in the previous cases, yielding the temperature-dependent residuals presented in Fig.~\ref{fig:residuals_panel}(B). The 1D-CNN exhibits excellent and uniform performance across this restricted range, reflected by an MAE of \SI{0.21}{\celsius}, an RMSE of \SI{0.26}{\celsius}, and a maximum error of \SI{0.82}{\celsius}. These values represent an approximate \SI{80}{\percent} reduction across all three error metrics when compared to the broader cryogenic case studied in Section~\ref{subsec:cryo}. This demonstrates that the machine learning approach is capable of sub-degree accuracy when calibrated over a narrow thermal window with adequately fine intervals. Such high-resolution capabilities are particularly valuable for biological applications, which typically require this level of thermometric precision in environments near room or physiological temperatures.

\subsection{Different Pump Power Densities}

Another relevant question for practical applications is how the 1D-CNN responds to spectra acquired under varying excitation laser power densities. It has been experimentally observed that the spectral morphology can change as a function of excitation power density~\cite{bujjamer2020luminescent,maestro2025intrinsic}. Developing a thermometer independent of excitation intensity is highly desirable for several reasons. For instance, excitation lasers may exhibit power drifts over time, or in dynamic systems, particles might move across the spatial profile of the laser beam, thereby experiencing fluctuations in the effective excitation intensity. To evaluate the thermometer's performance under such circumstances, three power densities were considered: \SI{5}{\watt\per\centi\meter\squared}, \SI{40}{\watt\per\centi\meter\squared}, and \SI{70}{\watt\per\centi\meter\squared}. The \SI{70}{\watt\per\centi\meter\squared} spectra are exactly the same dataset previously introduced in Section~\ref{subsec:cryo}, and were merged with spectra from two analogous thermal sweeps acquired at \SI{40}{\watt\per\centi\meter\squared} and \SI{5}{\watt\per\centi\meter\squared}. For these newly introduced measurements, approximately 250 spectra were acquired at each temperature setpoint.

The temperature-dependent residuals for the model's predictions in this scenario, trained and tested as in the previous cases, are presented in Fig.~\ref{fig:residuals_panel}(C). The observed performance closely mirrors that of previous evaluations with analogous thermal sweep ranges, demonstrating consistent and uniform accuracy across the investigated temperatures. Specifically, this calibration yielded an MAE of \SI{1.09}{\celsius}, an RMSE of \SI{1.45}{\celsius}, and a maximum prediction error of \SI{7.93}{\celsius}. The maximum error was the only metric to exhibit a considerable increase, rising by \SI{70}{\percent} compared to the baseline cryogenic case detailed in Section~\ref{subsec:cryo}. Nevertheless, the remaining metrics increased only marginally, showing that a single model trained on spectra from several power densities provides an accurate calibration across that range. This requires the relevant power densities to be represented in the training set: when applied to spectra at a power density not included in training, the model's predictions still vary monotonically with temperature but show a systematic calibration offset (data not shown).

\subsection{Different Pump Wavelengths}

A practical thermometer may also need to operate at different pump wavelengths, which can arise from laser retuning, mode hopping, or differences between experimental platforms. In lanthanide-doped upconversion systems, variations in pump wavelength can modify absorption efficiency, the relative population pathways, and the resulting spectral shape. While the $L^2$ normalization used here removes global intensity changes, wavelength-dependent changes in spectral morphology can still challenge model generalization. As a final, exploratory case study, luminescence spectra of Sample A were acquired at five temperature setpoints ranging from \SI{-140}{\celsius} to \SI{20}{\celsius} in \SI{40}{\celsius} increments. At each setpoint, approximately 100 spectra were recorded under three distinct excitation wavelengths: \SI{971}{\nano\meter}, \SI{976}{\nano\meter}, and \SI{981}{\nano\meter}.

The model was trained and tested following the same methodology as in previous cases, albeit with a much sparser temperature dataset. The resulting temperature-dependent residuals for this scenario are presented in Fig.~\ref{fig:residuals_panel}(D). Although the thermometer's performance is noticeably lower than in previous evaluations, the adverse nature of this dataset must be considered. The model was trained and tested using spectra acquired at only five widely spaced temperature setpoints, so these metrics should be read with this coarse sampling in mind. Furthermore, the majority of these spectra were recorded at excitation wavelengths detuned from the particles' primary absorption peak, which resulted in a substantially weaker emission signal and a lower signal-to-noise ratio. 

Under these conditions, the model yielded an MAE of \SI{2.08}{\celsius}, an RMSE of \SI{2.63}{\celsius}, and a maximum error of \SI{7.94}{\celsius}. These results show that a single 1D-CNN can be calibrated simultaneously for several excitation wavelengths, provided that spectra at each wavelength are included in the training set and a measurable luminescence signal is maintained.

\section{Conclusion}
We have shown that a one-dimensional convolutional neural network trained on normalized emission spectra provides an accurate and robust thermometer for $\text{NaYF}_4\text{:Yb}^{3+}\text{,Er}^{3+}$ upconversion micro- and nanoparticles from \SI{-150}{\celsius} to \SI{105}{\celsius}. In the two scenarios where both methods were compared directly, the 1D-CNN outperformed the conventional LIR, particularly at cryogenic temperatures, where the emission from the $^2\text{H}_{11/2}$ level approaches the detector noise floor, and under deliberate optical misalignment. By exploiting the full spectral morphology, including Stark-component features that remain temperature dependent where the LIR loses its sensitivity, the network keeps its temperature uncertainty nearly constant over the whole range and reaches this accuracy from single spectra, whereas the LIR would require averaging of many spectra. We further showed that the thermometer can be extended to a span exceeding \SI{250}{\celsius} at a modest cost in accuracy, reaches sub-degree accuracy over a narrow window, and that a single model can be calibrated across different excitation conditions (pump power and wavelength).

These conditions were reproduced here with particle ensembles to mimic the weak and fluctuating signals encountered with single particles and optically trapped particles. The natural next step is to apply this approach to single particles, both on substrates and in optical traps, where it could substantially shorten the integration times currently required for single-particle thermometry and help account for particle-to-particle variations of the emission spectra.

\begin{acknowledgement}
This work was supported by the Office of Naval Research Global (ONR Global) under Award No.~N62909-25-1-2030 and by Universidad de Buenos Aires under Grants No.~UBACyT 2023 (20020220400119BA) and No.~UBACyT 2026 (20020250300102BA). We thank David Sartor Maurino and Marco Garc\'{\i}a Novarini for upgrades of the cryostat, and Carlos Acha for providing the cryostat.
\end{acknowledgement}

\bibliography{biblio}

\end{document}